\documentclass[11pt,a4paper]{article} 
\usepackage{jheppub1}
\usepackage[english]{babel}
\usepackage{graphicx}% Include figure files
\usepackage{dcolumn}% Align table columns on decimal point
\usepackage{bm}% bold math
\usepackage{verbatim}
\usepackage{mathrsfs}
\usepackage{color}
\newcommand{\ket}[1]{\left| {#1} \right\rangle}
\newcommand{\bra}[1]{\left\langle {#1} \right|}

\newcommand{\proj}[2]{\left| {#1} \right\rangle\!\left\langle {#2} \right|}

\newcommand{\tr}{\operatorname{Tr}}

\newcommand{\eq}[1]{(\ref{#1})}

\newcommand{\up}{\uparrow}
\newcommand{\down}{\downarrow}
\newcommand{\qr}{q_\text{R}}

\def\slashchar#1{\setbox0=\hbox{$#1$} % set a box for #1
\dimen0=\wd0 % and get its size
\setbox1=\hbox{/} \dimen1=\wd1 % get size of /
\ifdim\dimen0>\dimen1 % #1 is bigger
\rlap{\hbox to \dimen0{\hfil/\hfil}} % so center / in box
#1 % and print #1
\else % / is bigger
\rlap{\hbox to \dimen1{\hfil$#1$\hfil}} % so center #1
/ % and print /
\fi}

\title{The entangling side of the Unruh-Hawking effect}
\author{Miguel Montero}
\affiliation{Instituto de F\'{i}sica Fundamental, CSIC, Serrano 113-B, 28006 Madrid, Spain}
\author{Eduardo Mart\'{i}n-Mart\'{i}nez}

\arxivnumber{1011.6540}

\emailAdd{montero89@gmail.com} \emailAdd{emmfis@gmail.com}  

\abstract{
We show that the Unruh effect can create net quantum entanglement between inertial and accelerated observers depending on the choice of the inertial state. This striking result banishes the extended belief that the Unruh effect can only destroy entanglement and furthermore provides a new and unexpected source for finding experimental evidence of the Unruh and Hawking effects.
}

\begin{document}
\toccontinuoustrue
\maketitle

\section{Introduction}

The influence of the so-called Unruh and Hawking effects \cite{Hawking,Hawking2,Hawking3,Hawking4} on quantum entanglement has been subject of many studies in the field of relativistic quantum information \cite{Alicefalls,Alicefalls2,AlsingSchul,AlsingSchul2,Edu4,Edu5,Bradler,Bradler2,Bradler3,Bradler4,Bradler5,Edu9}. In all these previous studies it was shown how starting with entangled states from an inertial perspective, we end up with a less entangled state when one of the observers is non-inertial. In this letter we show an unexpected outcome of these Unruh and Hawking effects: the appearance of entanglement when one of the observers of a bipartite system undergoes a constant acceleration.

For scalar fields \cite{Alicefalls,Alicefalls2,Edu4,Edu5} it has been shown that an inertial maximally entangled state loses entanglement from the perspective of a non-inertial observer as he accelerates. In this case the entanglement vanishes as the acceleration is increased. In the fermionic case the situation is more complex: the entanglement is degraded down to a finite limit \cite{AlsingSchul,AlsingSchul2,Edu4,Edu5}. In any case no net entanglement generation has ever been reported to happen due to the mere action of the Unruh effect. As a matter of fact, since the Unruh effect consists on the observation of a thermal distribution when an accelerated observer looks at the field vacuum, many times in relativistic quantum information this effect has been regarded as a source of decoherence comparable to a non-zero temperature environment.

We prove in this letter that there are some states, shared by two observers, whose degree of entanglement increases as one of the observers accelerates.  Note that this entanglement generation is radically different from the well-known entanglement creation reported elsewhere \cite{Hawking,Hawking2,Hawking3,Hawking4}. Namely, the entanglement between the two causally disconnected wedges of spacetime that arises when we express a separable state in the Rindler basis. Since classical communication between these two regions is forbidden, such entanglement could not be ever usefully measured to perform quantum information tasks. Our setting, however, considers an inertial observer and an accelerated one that can, in principle, signal one another.  The phenomenon is thoroughly studied here for Grassman scalar (a scalar field on which anticommutation relations are imposed) and bosonic scalar fields. 

This entanglement amplification is very promising in order to detect quantum effects due to acceleration (and therefore gravity). Entanglement, unlike other phenomena (such as thermal noise) does not admit a classical description. Thus, its observation would account for a pure quantum origin of the aforementioned effects. 

On the other hand, entanglement is very sensitive to any interaction with the environment, which tends to degrade it. This made it very difficult for any experiment relying on entanglement degradation \cite{Alicefalls,Alicefalls2} to find evidence for these effects. By the same token, experiments studying entanglement creation are free from these flaws: If a small amount of entanglement is created, no matter how damped by decoherence it may be, the only possible origin is an acceleration-induced quantum effect. The entanglement amplification phenomenon provides a novel way to distinguish genuine quantum effects of gravity from classically induced ones, something worth considering when trying to detect the Unruh and Hawking effects in analog gravity set-ups \cite{garay,garay2}.

The main reason why this phenomenon has gone unnoticed so far is the reliance in the so-called single mode approximation (SMA) \cite{Alsingtelep,Alsingtelep2} that many previous works assumed (\cite{Alicefalls,Alicefalls2,AlsingSchul,AlsingSchul2,Edu4,Edu5,Bradler,Bradler2,Bradler3,Bradler4,Bradler5} and many others). Such approximation consists on assumptions that are not generally true about the form of the change between two Fock bases, one built from monochromatic solutions of the Klein-Gordon or Dirac equation in Minkowski coordinates and other built from monochromatic solutions of these equations in Rindler coordinates. The proper way to work with monochromatic Rindler modes was shown in \cite{Edu6} and more detailedly in \cite{Edu9}, work that we will follow through all the letter. 

One could ask why entanglement seems to be created in this case. Observing an entangled state of the field from the perspective of an accelerated observer implies two processes: 1) a generation of entanglement due to the Bogoliubov relationships implied in the change of basis that was shadowed under the SMA and 2) an erasure of correlations due to the tracing over one of the Rindler regions. We will see that beyond the single mode approximation these competing trends explain why for a certain acceleration the amplification of entanglement is maximal. In previous works under the SMA it was simply not possible to see these two mechanisms in action.

\begin{figure}[hbtp] \centering
\includegraphics[width=.70\textwidth]{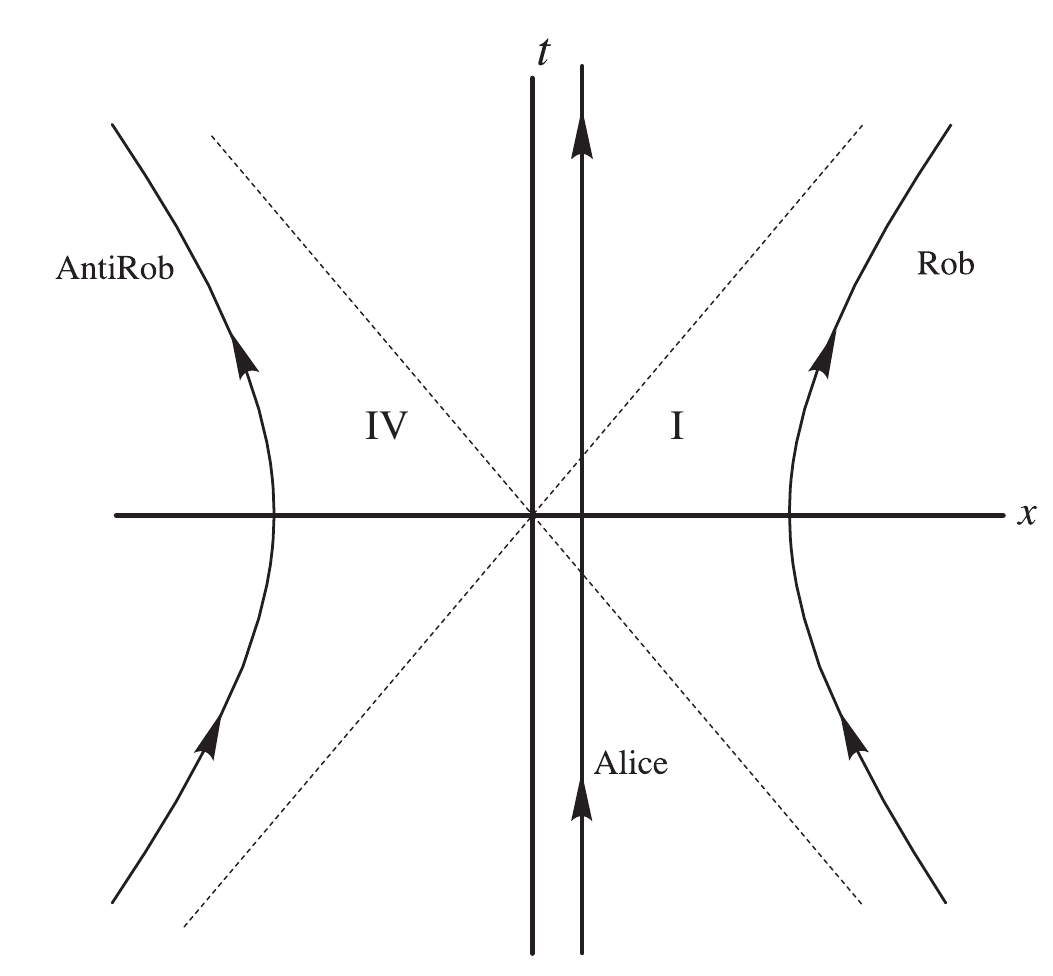}
\caption{Minkowski spacetime diagram showing the world lines of an inertial observer Alice, and one uniformly accelerated observer that can be moving in either the Rindler wedge I or IV which are causally disconnected from each other.}
\label{rind}
\end{figure}

\section{The setting}

Let us consider a system composed of an inertial observer, Alice, who watches an inertial mode of a quantum field (either a Grassman or bosonic scalar  field) and a uniformly accelerated Rindler observer either in region I or IV of Rindler spacetime. To follow previous literature we will call this observer  Rob if he is in region I and AntiRob if he is in region IV as illustrated in fig.\ref{rind}. Rob (or AntiRob) watches an Unruh mode \cite{Hawking,Hawking2,Hawking3,Hawking4} of the quantum field, which is entangled with Alice's. Unruh modes posses the peculiarity of having a sharp Rindler frequency. In other words, they correspond to monochromatic solutions of the field equation in Rindler coordinates, while an analytic continuation argument shows that they also are purely positive frequency linear combinations of Minkowski modes. This trivially implies that the vacuum state is the same both for Minkowski and Unruh modes. Furthermore, there are two kinds of Unruh modes \cite{Edu9}. Consideration of both kinds of modes is necessary for a complete description of an arbitrary solution to the field equation \cite{Hawking,Hawking2,Hawking3,Hawking4,Edu9}.

 We will only consider Unruh modes of a given Rindler frequency $\omega$ as seen by Rob or AntiRob (who move with proper acceleration $a$) but which are arbitrary superpositions of both kinds of Unruh modes mentioned above, which we label by the letters L and R, following the convenient notation in \cite{Edu9}: 
 \begin{align}C_\omega=q_\text{L}C_{\omega,\text{L}}+\qr C_{\omega,\text{R}}\label{umodes},\end{align}
where $\vert q_\text{L}\vert^2+\vert\qr\vert^2=1$, $\qr \ge q_\text{L}$ and the appropriate expressions for the operators in \eqref{umodes} are \cite{Edu9}
\begin{eqnarray}\label{bogoboson}
 C_{\omega,\text{R}}&=&\cosh r_\omega\, a_{\omega,\text{I}} - \sinh r_\omega\, a^\dagger_{\omega,\text{IV}},\\*
 C_{\omega,\text{L}}&=&\cosh r_\omega\, a_{\omega,\text{IV}} - \sinh r_\omega\, a^\dagger_{\omega,\text{I}}, \end{eqnarray}
where $a,a^\dagger$ are Rindler particle operators for the scalar field in each region, and
\begin{eqnarray}\label{Unruhop}
\nonumber C_{\omega,\text{\text{R}}}&=&\left(\cos r_k\, c_{\omega,\text{I}}-\sin r_\omega\, d^\dagger_{\omega,\text{IV}}\right),\\*
C_{\omega,\text{\text{L}}}&=&\left(\cos r_\omega\, c_{\omega,\text{IV}}-\sin r_\omega\, d^\dagger_{\omega,\text{I}}\right)
\end{eqnarray}
where $c,c^\dagger$ and $d,d^\dagger$ are respectively Rindler particle and antiparticle operators for the Grassman case. In this paper we will analyse the entanglement behaviour of the family of bipartite states 
\begin{align}\label{geGras} \ket{\Psi}\nonumber &= P \ket{0}_\text{A} \left[\alpha\ket{1_\omega} +\sqrt{1-\alpha^2}\ket{0_\omega}\right] \\
 &+ \sqrt{1-P^2}\ket{1}_\text{A} \left[\beta\ket{1_\omega}+ \sqrt{1-\beta^2}\ket{0_\omega}\right]
.\end{align}
Here, the subscript `A' refers to Alice's inertial mode, and $\ket{1_\omega}=C^\dagger_\omega\ket{0}$ is the Unruh particle excitation. All these states have an implicit dependence on Rob's acceleration $a$ when expressed in the Rindler basis through a parameter $r_\omega$ defined by $\tan r_\omega=e^{-\pi c\,\omega/a}$ in the fermionic case, and $\tanh r_\omega=e^{-\pi c\,\omega/a}$ in the bosonic case.

As usual \cite{Alicefalls,Alicefalls2,AlsingSchul,AlsingSchul2}, we transform the state \eqref{geGras} into the Rindler basis following the same conventions as in \cite{Edu9}. This is done because Rindler modes are those natural to an accelerated observers much in the same way as Minkowski modes are natural to inertial observers. Although the system is obviously bipartite, shifting to the Rindler basis for the second qubit the mathematical description of the system \cite{Alicefalls,Alicefalls2,AlsingSchul,AlsingSchul2,Edu4,Edu5} admits a straightforward tripartition: Minkowskian modes, Rindler region I modes, and Rindler region IV  modes. 

The density matrix for the state, which includes modes on both wedges of the spacetime along with Minkowskian modes, is built from \eqref{geGras}. Namely, $\rho^{\text{AR}\bar{\text{R}}}=\proj{\Psi}{\Psi}$.

As shown in figure \ref{rind}, an accelerated observer in region I is causally disconnected from region IV (and vice-versa). For this reason when we consider the bipartite system Alice-Rob we need to trace over the modes that only have support in region IV and from which Rob is causally disconnected. Equivalently, we would have to trace over modes in region I if we consider that the accelerated observer is in region IV. The density matrices for the bipartite systems AR and $\text{A}{\bar{\text{R}}}$ are  (See \cite{Alicefalls,Alicefalls2,AlsingSchul,AlsingSchul2,Edu4,Edu5,Edu9})
\begin{eqnarray}
\label{AR2}\rho^{\text{AR}}&=&\tr_{\text{IV}}\rho^{\text{AR}\bar{\text{R}}}_d=\sum_{n} \bra{n}_{\text{IV}}\rho^{\text{AR}\bar{\text{R}}}\ket{n}_{\text{IV}}\\*
\label{AAR2}\rho^{\text{A}\bar{\text{R}}}&=&\tr_{\text{I}}\rho^{\text{AR}\bar{\text{R}}}=\sum_{n} \bra{n}_{\text{I}}\rho^{\text{AR}\bar{\text{R}}}\ket{n}_{\text{I}}.
\end{eqnarray}

 We are interested in quantifying the entanglement between Alice and Rob, (or between Alice and AntiRob). To do so, we trace out AntiRob's or Rob's degrees of freedom respectively, and compute the negativities \cite{Negat} $\mathcal{N}_{\text{AR}}$ and $\mathcal{N}_{\text{A}\bar{\text{R}}}$ of the resulting reduced states as it is commonplace in the field \cite{Alicefalls,Alicefalls2,AlsingSchul,AlsingSchul2,Edu9}.

Any result for the Grassman field can be automatically carried over to the Dirac field provided that we focus only on those Dirac states with Grassman analog (namely, a state which can be formally converted into a Grassman state by the replacement $\ket{\up}\ \rightarrow \ \ket{1}$, or $\ket{\down}\ \rightarrow\ \ket{1}$). This is due to the fact that for fixed spin z-component the Dirac field has the same algebraic structure as the Grassman field \cite{AlsingSchul,AlsingSchul2,Edu4,Edu5}. The converse holds as well, as long as we consider only states with Grassman analog. 

\section{Entanglement amplification}

When one studies entanglement of fermionic fields in non-inertial frames a somewhat surprising result appears: some entanglement survives in the infinite acceleration limit \cite{AlsingSchul,AlsingSchul2}. A naive reasoning might conclude that since the Unruh effect acts as a thermal bath with $T\propto a$, all the entanglement should be washed out by an infinite temperature thermal decoherence. This argument is flawed because the Unruh thermal state is derived for the Minkowski vacuum state, not for states containing excitations. Yet, previous works under the SMA found that entanglement in the AR bipartition was a monotone decreasing function of $r_\omega$. So it still seemed true that acceleration tends to degrade entanglement. When studying the general states \eqref{geGras} these trends would be expected to continue holding in principle.

\begin{figure}[hbtp] \centering
\includegraphics[width=.70\textwidth]{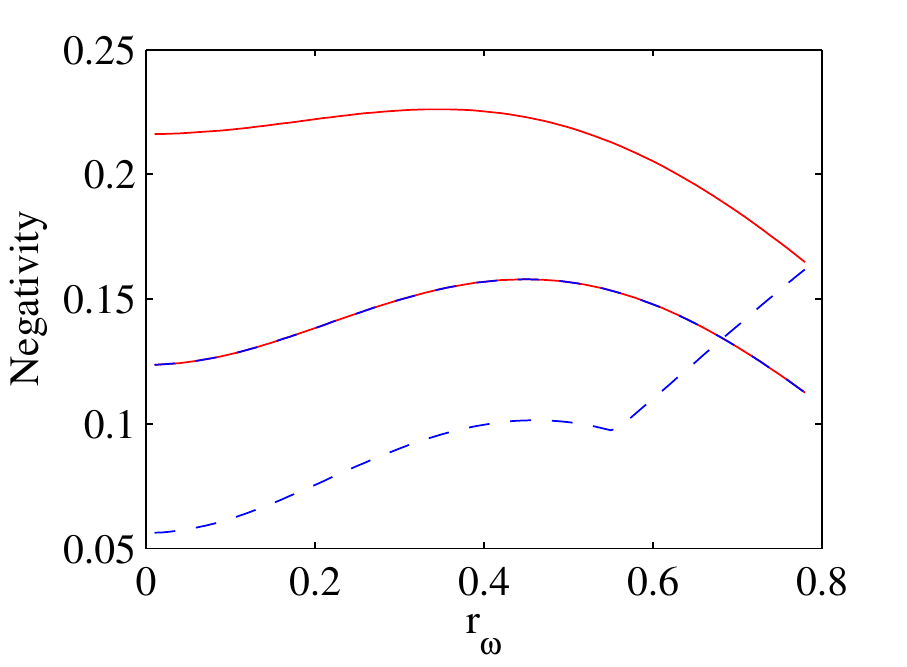}
\caption{Entanglement creation for the Grassman scalar field. Negativity for $AR$ (red continuous) and $A\bar R$ (blue dashed) as a function of $r_\omega$ for $\qr=0.85$ and $\qr=1/\sqrt{2}$ (where both contributions are equal). The state considered is of the general form \eqref{geGras}, with $P=0.4$, $\alpha=0$, $\beta=1$. }
\label{ent}
\end{figure}

Instead, for $\qr\neq1$ and a rather simple choice of parameters (for instance, $P=0.4$, $\alpha=0$, $\beta=1$) the surprise appears: there can be entanglement amplification due to acceleration, as seen in fig. \ref{ent}. Furthermore, this amplification becomes more evident as $\qr$ approaches the extremal value $\qr=1/\sqrt{2}$.  AR and $\text{A}\bar{\text{R}}$ behave the same way in this case since the symmetry between regions I and IV is not explicitly broken (see eq. \eqref{umodes} and ref. \cite{Edu9}). As $\qr$ tends to 1 (limit referred in previous works as the SMA), the effect vanishes. 

More importantly, it is also possible to obtain high entanglement amplification considering almost separable states. In some of these cases the negativity is a monotone function of $r_\omega$ (as for instance in \eqref{geGras} when $P=0.1$, $\alpha=0.8695$, $\beta=0.909$). Therefore, there are states for which the Unruh effect does exactly the opposite of what was expected. That is to say, entanglement is monotonically created rather than being monotonically destroyed.

\begin{figure}[hbtp] \centering
\includegraphics[width=.70\textwidth]{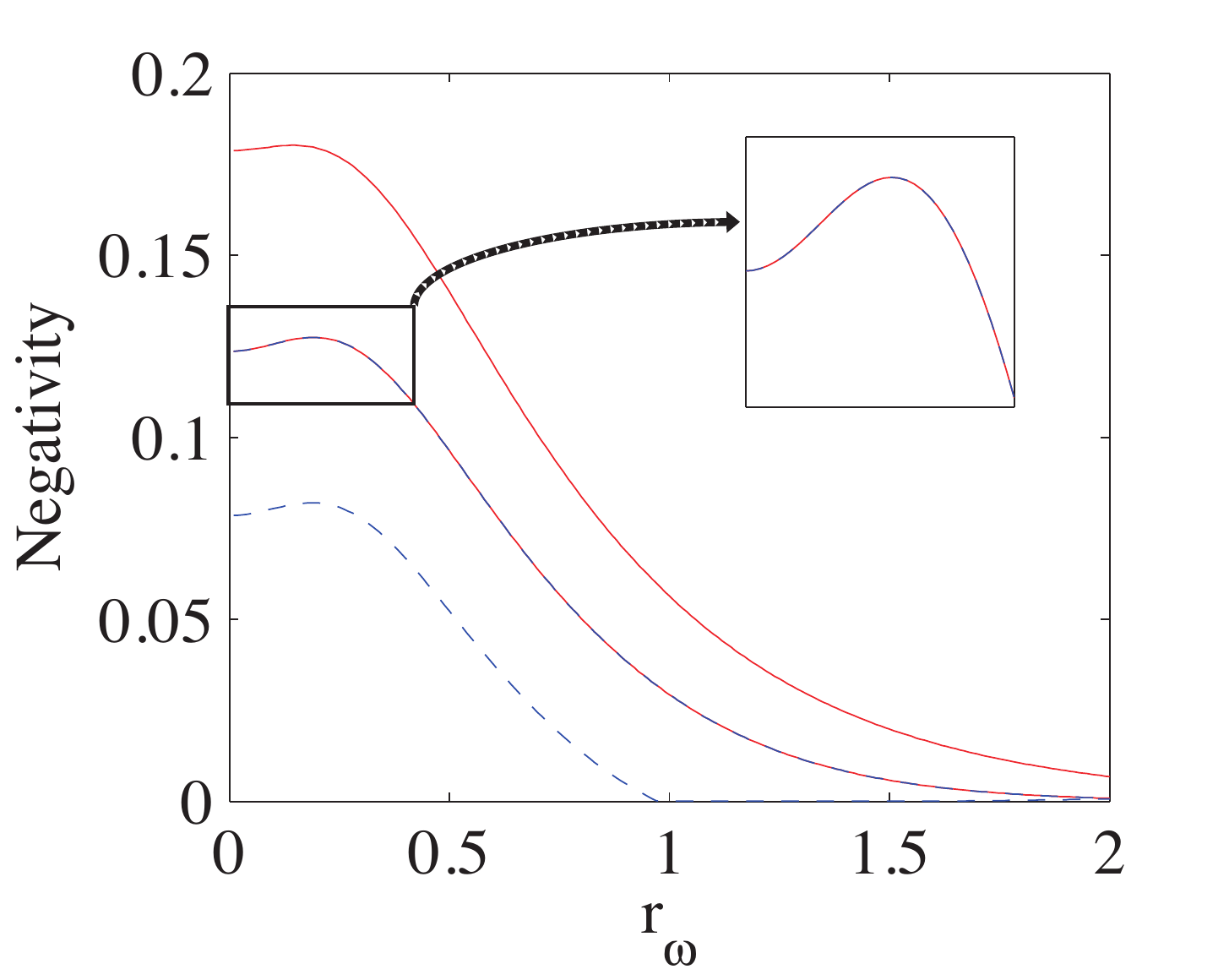}
\caption{Entanglement creation for the bosonic scalar field. Negativity for $AR$ (red continuous) and $A\bar R$ (blue dashed) as a function of $r_\omega$ for $\qr=0.85$ and $\qr=1/\sqrt{2}$ (where both contributions are equal). The state considered is of the general form \eqref{geGras}, with $P=0.4$, $\alpha=0$, $\beta=1$.}
\label{ent2}
\end{figure}

The phenomenon of entanglement creation also shows up for the bosonic scalar field, much in the same way as it did in the Grassman case. The main difference is that  in the bosonic case entanglement is bound to vanish in the infinite acceleration limit, in concordance with previous results \cite{Edu9}. Therefore, entanglement can be created only for a finite range of accelerations, as shown in fig.\ref{ent2}. For $\qr=1/\sqrt{2}$, Alice-Rob negativity attains a maximum of $0.127$ at $r_\omega=0.191$. This is  3.1 \% above inertial level. The definition of $r_\omega$ in bosonic fields differs from that of fermionic fields in the replacement $\tan\rightarrow\tanh$, and therefore the allowed range of $r_\omega$ is $[0,\infty)$ rather than $[0,\frac{\pi}{4})$. Considering frequencies of order $\approx 1$  GHz, which correspond to reasonable experimental possibilities, \cite{Adessada} the acceleration corresponding to this value of $r_\omega$ is $a\approx 10^{17}g$, much closer to experimental feasibility than previous proposals \cite{ChenTaj} which suggested accelerations of $\approx 10^{25}g$.
The appearance of entanglement creation in the bosonic case shows that this is a truly universal phenomenon, independent of field statistics.

In order to study the experimental implications of this phenomenon, let us introduce a specific scenario. Consider the family of states \eq{geGras} for fixed $P$ and $\beta \gtrsim 0.2$. This family happens to show an unbounded relative increase of entanglement  as one approaches the limit $\alpha\rightarrow \beta$ (separable limit). This means that  there are states for which an arbitrary small acceleration produces an arbitrary large relative increase in negativity. The same happens as  we approach a separable state taking $P\rightarrow0$ for certain values of $\alpha$ and $\beta$. This behaviour is quite general and appears for both fermionic and bosonic fields. However the more relative entanglement increase (signal-to-background ratio) we want to achieve, the more separable the inertial states should be. Dealing with quasi-separable states would be the experimental challenge to detect the Unruh effect by means of these techniques.

\begin{figure}[hbtp] \centering
\includegraphics[width=.70\textwidth]{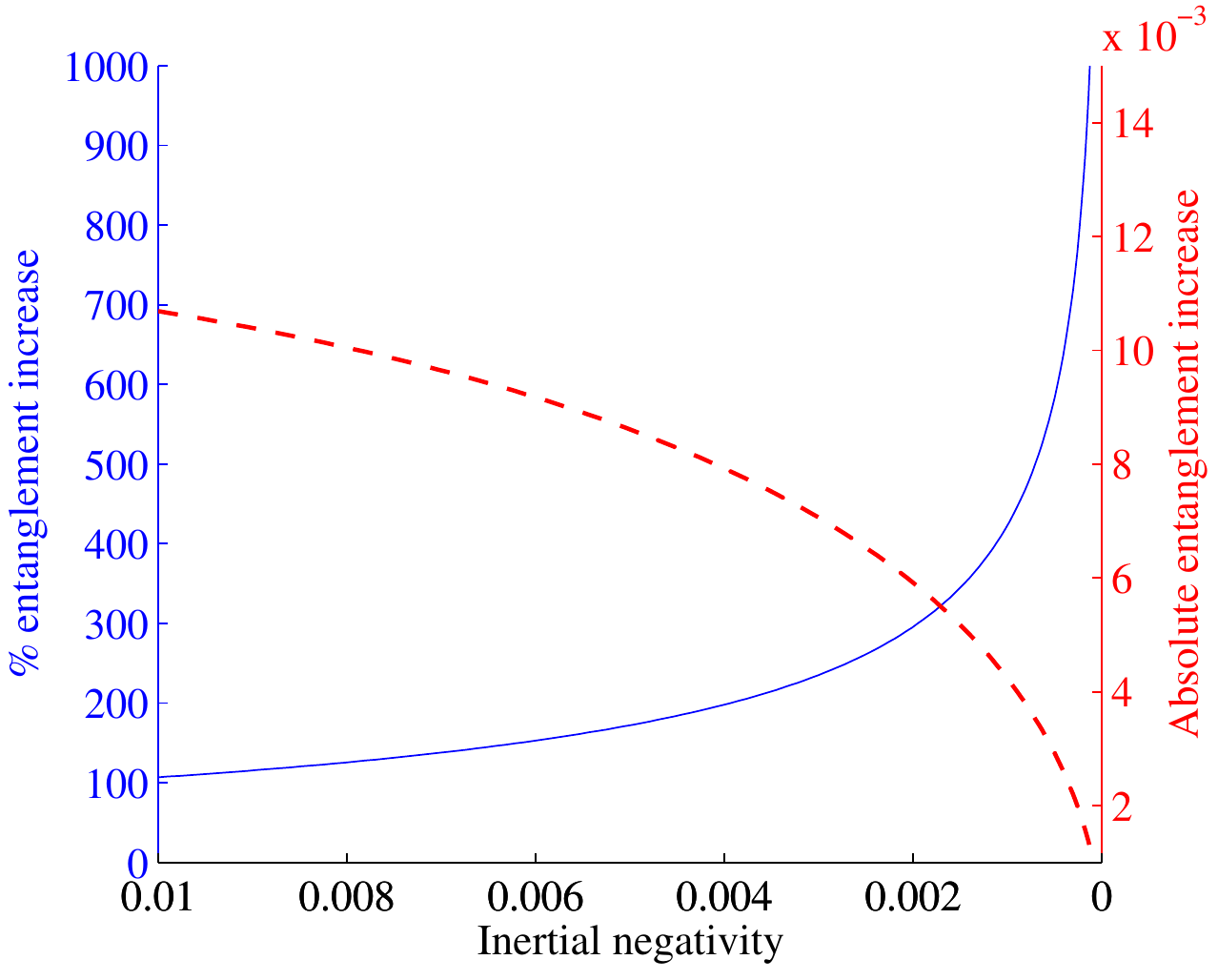}
\caption{Relative (blue continuous) and absolute (red dashed) entanglement creation  as a function of the inertial negativity for a Grassman field  state of the family \eqref{geGras} with $P=0.4$, $\beta=0.8$ and different values of $\alpha$ for $\qr=1/\sqrt{2}$ and  $r_\omega=0.15$ (accelerations from $a\approx 5\cdot10^{13}g $ to $5\cdot10^{16} g$ for frequencies from $1 \text{ MHz}$ to $1 \text{ GHz}$). Notice an unbounded growth of the `signal-to-background' ratio.}
\label{unbo}
\end{figure}

Any such experiment would be naturally interested in negativity behaviour in the vicinity of $r_\omega=0$, easier to obtain in laboratory conditions. This means that in order to maximise experimental feasibility we are interested in states whose negativity shows a quick growth for small $r_\omega$. We study the relative increase of AR negativity with respect to its inertial value for the family of states \eqref{geGras} at fixed $r_\omega$. As an example we choose  $r_\omega=0.15$ which corresponds to accelerations from $a\approx 5\cdot10^{13}g $ to $5\cdot10^{16} g$ for frequencies from $1 \text{ MHz}$ to $1 \text{ GHz}$. This unbounded entanglement creation can be seen in fig. \ref{unbo}.

We obtain huge `signal-to-background' ratios and the better the negativity can be experimentally determined, the bigger this ratio can become. If we relied on entanglement degradation to detect the Unruh effect \cite{Alicefalls,Alicefalls2}, the percental change in negativity would be bounded by 100~\%. With the plethora of new states presented in this work, this relative change can be made arbitrarily high.
 
The same analysis carried out for Unruh modes can be repeated if we consider that the excitations $\ket{1_\omega}$ are localised Gaussian wavepackets from the inertial perspective. As detailedly shown in section IV of \cite{Edu9}, Gaussian wavepackets of Minkowski modes transform into Gaussian wavepackets of Rindler modes. We can consider at the same time peaked wavepackets in the Minkowskian basis and in the Rindler basis such that the analysis would be completely analogous to the monochromatic case. In this case different choices of $q_\text{R}$ and $q_\text{L}$ represent different spatial localisation of the Gaussian wavepackets. This means that in principle we can define two local bases (for Alice and Rob) in which this entanglement creation phenomenon is present.

Additional mathematical details that may help follow the computational scaffolding underlying the results presented in this letter are included as additional material.

\section{Conclusions}

We have shown that the Unruh effect can not only degrade quantum entanglement shared by an inertial and an accelerated observer but also amplify it, banishing previous fundamental misconceptions such as the belief that the Unruh and Hawking effects are sources of entanglement degradation. We have demonstrated that there are families of states whose entanglement can be increased by an arbitrarily high relative factor.

Furthermore, with these results we move the experimental difficulties from generating and sustaining high accelerations to being able to prepare and measure quasi-separable entangled states, hence, we are presenting a new way to detect the Unruh effect. As a matter of fact, our results are independent of the specific implementation to detect the entanglement magnification,  hence they can be exported to a huge variety of experimental set-ups as for instance analog gravity settings.

Although only scalar and Grassman scalar fields have been analysed here, a similar phenomenon also happens in higher spins such as Dirac and electromagnetic fields whose study constitutes itself an interesting piece of work that will be reported elsewhere \cite{migsp}.

All these results can be readily exported to a setting consisting of an observer hovering at certain distance close to the event horizon of an Schwarzschild black hole, following \cite{Edu6}, and therefore the same conclusions drawn for the Unruh effect are also valid for the Hawking effect.

\section{Acknowledgements}
We thank J. Louko, A. Dragan, I. Fuentes and T. Ralph for interesting discussions and comments during the RQI4 workshop held in Brisbane (Australia). We also thank J. Le\'on for his useful comments. E. M-M was supported by a CSIC JAE-PREDOC2007 Grant and by the Spanish MICINN Project FIS2008-05705/FIS and the QUITEMAD consortium.


\begin{thebibliography}{25}
\expandafter\ifx\csname natexlab\endcsname\relax\def\natexlab#1{#1}\fi
\expandafter\ifx\csname bibnamefont\endcsname\relax
  \def\bibnamefont#1{#1}\fi
\expandafter\ifx\csname bibfnamefont\endcsname\relax
  \def\bibfnamefont#1{#1}\fi
\expandafter\ifx\csname citenamefont\endcsname\relax
  \def\citenamefont#1{#1}\fi
\expandafter\ifx\csname url\endcsname\relax
  \def\url#1{\texttt{#1}}\fi
\expandafter\ifx\csname urlprefix\endcsname\relax\def\urlprefix{URL }\fi
\providecommand{\bibinfo}[2]{#2}
\providecommand{\eprint}[2][]{\url{#2}}


\bibitem[{\citenamefont{Hawking}(1974)}]{Hawking}
\bibinfo{author}{\bibfnamefont{S.~W.} \bibnamefont{Hawking}},
  \bibinfo{journal}{Nature} \textbf{\bibinfo{volume}{248}}, \bibinfo{pages}{30}
  (\bibinfo{year}{1974}). 
  
 \bibitem[{\citenamefont{Unruh}(1976)}]{Hawking2}
  \bibinfo{author}{\bibfnamefont{W.~G.} \bibnamefont{Unruh}},
  \bibinfo{journal}{Phys. Rev. D} \textbf{\bibinfo{volume}{14}},
  \bibinfo{pages}{870} (\bibinfo{year}{1976}). 
  
  \bibitem[{\citenamefont{Takagi}(1986)}]{Hawking3}
  \bibinfo{author}{\bibfnamefont{S.}~\bibnamefont{Takagi}},
  \bibinfo{journal}{Prog. Theor. Phys. Suppl.} \textbf{\bibinfo{volume}{88}},
  \bibinfo{pages}{1} (\bibinfo{year}{1986}).
  
  \bibitem[{\citenamefont{Langlois}(2004)}]{Hawking4}
  \bibinfo{author}{\bibfnamefont{P.}~\bibnamefont{Langlois}},
  \bibinfo{journal}{Phys. Rev. D} \textbf{\bibinfo{volume}{70}},
  \bibinfo{pages}{104008} (\bibinfo{year}{2004}).


\bibitem[{\citenamefont{Fuentes-Schuller and Mann}(2005)}]{Alicefalls}
\bibinfo{author}{\bibfnamefont{I.}~\bibnamefont{Fuentes-Schuller}}
  \bibnamefont{and} \bibinfo{author}{\bibfnamefont{R.~B.} \bibnamefont{Mann}},
  \bibinfo{journal}{Phys. Rev. Lett.} \textbf{\bibinfo{volume}{95}},
  \bibinfo{pages}{120404} (\bibinfo{year}{2005}).
  
 \bibitem[{\citenamefont{J. Le?on and E. Mart\' in-Mart\' inez}(2010)}]{Alicefalls2} 
   \bibinfo{author}{\bibfnamefont{E.}~\bibnamefont{Mart\'\i{}n-Mart\'\i{}nez}}
  \bibnamefont{and} \bibinfo{author}{\bibfnamefont{J.}~\bibnamefont{Le\'on}},
  \bibinfo{journal}{Phys. Rev. A} \textbf{\bibinfo{volume}{81}},
  \bibinfo{pages}{052305} (\bibinfo{year}{2010}{\natexlab{b}}).

\bibitem[{\citenamefont{Alsing et~al.}(2006)\citenamefont{Alsing,
  Fuentes-Schuller, Mann, and Tessier}}]{AlsingSchul}
\bibinfo{author}{\bibfnamefont{P.~M.} \bibnamefont{Alsing}},
  \bibinfo{author}{\bibfnamefont{I.}~\bibnamefont{Fuentes-Schuller}},
  \bibinfo{author}{\bibfnamefont{R.~B.} \bibnamefont{Mann}}, \bibnamefont{and}
  \bibinfo{author}{\bibfnamefont{T.~E.} \bibnamefont{Tessier}},
  \bibinfo{journal}{Phys. Rev. A} \textbf{\bibinfo{volume}{74}},
  \bibinfo{pages}{032326} (\bibinfo{year}{2006}).
 
 \bibitem[{\citenamefont{J. Le?on and E. Mart\' in-Mart\' inez}(2009)}]{AlsingSchul2} 
\bibinfo{author}{\bibfnamefont{J.}~\bibnamefont{Le\'on}} \bibnamefont{and}
  \bibinfo{author}{\bibfnamefont{E.}~\bibnamefont{Mart\'\i{}n-Mart\'\i{}nez}},
  \bibinfo{journal}{Phys. Rev. A} \textbf{\bibinfo{volume}{80}},
  \bibinfo{pages}{012314} (\bibinfo{year}{2009}).

\bibitem[{\citenamefont{Mart\'\i{}n-Mart\'\i{}nez and Le\'on}(2009)}]{Edu4}
\bibinfo{author}{\bibfnamefont{E.}~\bibnamefont{Mart\'\i{}n-Mart\'\i{}nez}}
  \bibnamefont{and} \bibinfo{author}{\bibfnamefont{J.}~\bibnamefont{Le\'on}},
    \bibinfo{journal}{Phys. Rev. A} \textbf{\bibinfo{volume}{80}},
  \bibinfo{pages}{042318} (\bibinfo{year}{2009}).


\bibitem[{\citenamefont{Mart\'\i{}n-Mart\'\i{}nez and Le\'on}(2009)}]{Edu5}
\bibinfo{author}{\bibfnamefont{E.}~\bibnamefont{Mart\'\i{}n-Mart\'\i{}nez}}
  \bibnamefont{and} \bibinfo{author}{\bibfnamefont{J.}~\bibnamefont{Le\'on}},
   \bibinfo{journal}{Phys. Rev. A} \textbf{\bibinfo{volume}{81}},
\bibinfo{pages}{032320} (\bibinfo{year}{2010}{\natexlab{a}}).


\bibitem[{\citenamefont{Br\'adler}(2007)}]{Bradler}
\bibinfo{author}{\bibfnamefont{K.}~\bibnamefont{Br\'adler}},
  \bibinfo{journal}{Phys. Rev. A} \textbf{\bibinfo{volume}{75}},
  \bibinfo{pages}{022311} (\bibinfo{year}{2007}).

\bibitem[{\citenamefont{X.-H. Ge and S. P. Kim}(2008)}]{Bradler2}
\bibinfo{author}{\bibfnamefont{X.-H.} \bibnamefont{Ge}} \bibnamefont{and}
  \bibinfo{author}{\bibfnamefont{S.~P.} \bibnamefont{Kim}},
  \bibinfo{journal}{Class. Quantum Grav.} \textbf{\bibinfo{volume}{25}},
  \bibinfo{pages}{075011} (\bibinfo{year}{2008}).

\bibitem[{\citenamefont{S. Moradi}(2009)}]{Bradler3}
\bibinfo{author}{\bibfnamefont{S.}~\bibnamefont{Moradi}},
  \bibinfo{journal}{Phys. Rev. A} \textbf{\bibinfo{volume}{79}},
  \bibinfo{pages}{064301} (\bibinfo{year}{2009}).

\bibitem[{\citenamefont{A. G. S. Landulfo and G. E. A. Matsas}(2009)}]{Bradler4}
\bibinfo{author}{\bibfnamefont{A.~G.~S.} \bibnamefont{Landulfo}}
  \bibnamefont{and} \bibinfo{author}{\bibfnamefont{G.~E.~A.}
  \bibnamefont{Matsas}}, \bibinfo{journal}{Phys. Rev. A}
  \textbf{\bibinfo{volume}{80}}, \bibinfo{pages}{032315}
  (\bibinfo{year}{2009}).

\bibitem[{\citenamefont{D. C. M. Ostapchuk and R. B. Mann}(2009)}]{Bradler5}
\bibinfo{author}{\bibfnamefont{D.~C.~M.} \bibnamefont{Ostapchuk}}
  \bibnamefont{and} \bibinfo{author}{\bibfnamefont{R.~B.} \bibnamefont{Mann}},
  \bibinfo{journal}{Phys. Rev. A} \textbf{\bibinfo{volume}{79}},
  \bibinfo{pages}{042333} (\bibinfo{year}{2009}).
  
\bibitem[{\citenamefont{Bruschi et~al.}(2010)\citenamefont{Bruschi, Louko,
  Mart\'\i{}n-Mart\'\i{}nez, Dragan, and Fuentes}}]{Edu9}
\bibinfo{author}{\bibfnamefont{D.~E.} \bibnamefont{Bruschi}},
  \bibinfo{author}{\bibfnamefont{J.}~\bibnamefont{Louko}},
  \bibinfo{author}{\bibfnamefont{E.}~\bibnamefont{Mart\'\i{}n-Mart\'\i{}nez}},
  \bibinfo{author}{\bibfnamefont{A.}~\bibnamefont{Dragan}}, \bibnamefont{and}
  \bibinfo{author}{\bibfnamefont{I.}~\bibnamefont{Fuentes}},
  \bibinfo{journal}{Phys. Rev. A} \textbf{\bibinfo{volume}{82}},
  \bibinfo{pages}{042332} (\bibinfo{year}{2010}). 
  
  

\bibitem[{\citenamefont{Alsing and Milburn}(2003)}]{Alsingtelep}
\bibinfo{author}{\bibfnamefont{P.~M.} \bibnamefont{Alsing}} \bibnamefont{and}
  \bibinfo{author}{\bibfnamefont{G.~J.} \bibnamefont{Milburn}},
  \bibinfo{journal}{Phys. Rev. Lett.} \textbf{\bibinfo{volume}{91}},
  \bibinfo{pages}{180404} (\bibinfo{year}{2003}).
  
  \bibitem[{\citenamefont{P. M. Alsing, D. McMahon, and G. J. Milburn}(2003)}]{Alsingtelep2}
   \bibinfo{author}{\bibfnamefont{P.~M.} \bibnamefont{Alsing}},
  \bibinfo{author}{\bibfnamefont{D.}~\bibnamefont{McMahon}}, \bibnamefont{and}
  \bibinfo{author}{\bibfnamefont{G.~J.} \bibnamefont{Milburn}},
  \bibinfo{journal}{J. Opt. B: Quantum Semiclass. Opt.}
  \textbf{\bibinfo{volume}{6}}, \bibinfo{pages}{S834} (\bibinfo{year}{2004}).
  
  
  
  \bibitem[{\citenamefont{Mart\'\i{}n-Mart\'\i{}nez
  et~al.}(2010)\citenamefont{Mart\'\i{}n-Mart\'\i{}nez, Garay, and
  Le\'on}}]{Edu6}
\bibinfo{author}{\bibfnamefont{E.}~\bibnamefont{Mart\'\i{}n-Mart\'\i{}nez}},
  \bibinfo{author}{\bibfnamefont{L.~J.} \bibnamefont{Garay}}, \bibnamefont{and}
  \bibinfo{author}{\bibfnamefont{J.}~\bibnamefont{Le\'on}},
  \bibinfo{journal}{Phys. Rev. D} \textbf{\bibinfo{volume}{82}},
  \bibinfo{pages}{064006} (\bibinfo{year}{2010}).



\bibitem[{\citenamefont{Vidal and Werner}(2002)}]{Negat}
\bibinfo{author}{\bibfnamefont{G.}~\bibnamefont{Vidal}} \bibnamefont{and}
  \bibinfo{author}{\bibfnamefont{R.~F.} \bibnamefont{Werner}},
  \bibinfo{journal}{Phys. Rev. A} \textbf{\bibinfo{volume}{65}},
  \bibinfo{pages}{032314} (\bibinfo{year}{2002}).

\bibitem[{\citenamefont{Aspachs et~al.}(2010)\citenamefont{Aspachs, Adesso, and
  Fuentes}}]{Adessada}
\bibinfo{author}{\bibfnamefont{M.}~\bibnamefont{Aspachs}},
  \bibinfo{author}{\bibfnamefont{G.}~\bibnamefont{Adesso}}, \bibnamefont{and}
  \bibinfo{author}{\bibfnamefont{I.}~\bibnamefont{Fuentes}},
  \bibinfo{journal}{Phys. Rev. Lett.} \textbf{\bibinfo{volume}{105}},
  \bibinfo{pages}{151301} (\bibinfo{year}{2010}).

\bibitem[{\citenamefont{Chen and Tajima}(1999)}]{ChenTaj}
\bibinfo{author}{\bibfnamefont{P.}~\bibnamefont{Chen}} \bibnamefont{and}
  \bibinfo{author}{\bibfnamefont{T.}~\bibnamefont{Tajima}},
  \bibinfo{journal}{Phys. Rev. Lett.} \textbf{\bibinfo{volume}{83}},
  \bibinfo{pages}{256} (\bibinfo{year}{1999}).
  
  

\bibitem[{\citenamefont{Garay et~al.}(2000)\citenamefont{Garay, Anglin, Cirac,
  and Zoller}}]{garay}
\bibinfo{author}{\bibfnamefont{L.~J.} \bibnamefont{Garay}},
  \bibinfo{author}{\bibfnamefont{J.~R.} \bibnamefont{Anglin}},
  \bibinfo{author}{\bibfnamefont{J.~I.} \bibnamefont{Cirac}}, \bibnamefont{and}
  \bibinfo{author}{\bibfnamefont{P.}~\bibnamefont{Zoller}},
  \bibinfo{journal}{Phys. Rev. Lett.} \textbf{\bibinfo{volume}{85}},
  \bibinfo{pages}{4643} (\bibinfo{year}{2000}).
 
 \bibitem[{\citenamefont{Horstmann et~al.}(2010)\citenamefont{B. Horstmann, B. Reznik, S. Fagnocchi, and J. I. Cirac}}]{garay2}
  \bibinfo{author}{\bibfnamefont{B.}~\bibnamefont{Horstmann}},
  \bibinfo{author}{\bibfnamefont{B.}~\bibnamefont{Reznik}},
  \bibinfo{author}{\bibfnamefont{S.}~\bibnamefont{Fagnocchi}},
  \bibnamefont{and} \bibinfo{author}{\bibfnamefont{J.~I.} \bibnamefont{Cirac}},
  \bibinfo{journal}{Phys. Rev. Lett.} \textbf{\bibinfo{volume}{104}},
  \bibinfo{pages}{250403} (\bibinfo{year}{2010}).

 \bibitem[{\citenamefont{Mons}(2010)\citenamefont{B. Horstmann, B. Reznik, S. Fagnocchi, and J. I. Cirac}}]{migsp}
  \bibinfo{author}{\bibfnamefont{M.}~\bibnamefont{Montero}},
  \bibnamefont{and} \bibinfo{author}{\bibfnamefont{E.} \bibnamefont{Mart\'in-Mart\'inez}},
  \bibinfo{journal}{to appear in Phys. Rev. A},
  \bibinfo{pages}{arXiv:1105.0894} (\bibinfo{year}{2011}).
  

%\bibitem[{\citenamefont{Bennett et~al.}(1996)\citenamefont{Bennett, Bernstein,
 % Popescu, and Schumacher}}]{Distill}
%\bibinfo{author}{\bibfnamefont{C.~H.} \bibnamefont{Bennett}},
 % \bibinfo{author}{\bibfnamefont{H.~J.} \bibnamefont{Bernstein}},
 % \bibinfo{author}{\bibfnamefont{S.}~\bibnamefont{Popescu}}, \bibnamefont{and}
 % \bibinfo{author}{\bibfnamefont{B.}~\bibnamefont{Schumacher}},
 % \bibinfo{journal}{Phys. Rev. A} \textbf{\bibinfo{volume}{53}},
 % \bibinfo{pages}{2046} (\bibinfo{year}{1996}).

\end{thebibliography}
\end{document}